\documentclass[aps,prb,twocolumn,longbibliography,superscriptaddress,article]{revtex4-1}
\usepackage{epsfig}
\usepackage{epstopdf}
\usepackage{amsmath}
\usepackage{amsfonts}
\usepackage{amssymb}
\usepackage{hyperref}
\usepackage{bm}
\usepackage{makecell}
\usepackage{rotating}
\usepackage{hyperref}
\usepackage{multirow}
\usepackage{graphicx}
\usepackage{booktabs}
\usepackage{siunitx}
\usepackage{makecell}
\usepackage{color, soul}

\usepackage{graphicx}
\usepackage{dcolumn}
\usepackage{bm}
\usepackage{color}

\usepackage{tikz,xcolor,hyperref}

\definecolor{lime}{HTML}{A6CE39}
\DeclareRobustCommand{\orcidicon}{%
	\begin{tikzpicture}
	\draw[lime, fill=lime] (0,0)
	circle [radius=0.16]
	node[white] {{\fontfamily{qag}\selectfont \tiny ID}};
	\draw[white, fill=white] (-0.0625,0.095)
	circle [radius=0.007];
	\end{tikzpicture}
	\hspace{-2mm}
}

\foreach \x in {A, ..., Z}{%
	\expandafter\xdef\csname orcid\x\endcsname{\noexpand\href{https://orcid.org/\csname orcidauthor\x\endcsname}{\noexpand\orcidicon}}
}

\begin{document}

\title{Quantum anomalous Hall conductivity in altermagnets under applied magnetic field}

\author{Meysam Bagheri Tagani\orcidA}
\email{mtagani@magtop.ifpan.edu.pl}
\affiliation{International Research Centre Magtop, Institute of Physics, Polish Academy of Sciences, Aleja Lotnik\'ow 32/46, PL-02668 Warsaw, Poland}
\affiliation{Department of Physics, University of Guilan, P. O. Box 41335-1914, Rasht, Iran}

\author{Amar Fakhredine\orcidF}
\affiliation{Institute of Physics, Polish Academy of Sciences, Aleja Lotnik\'ow 32/46, 02668 Warsaw, Poland}

\author{Carmine Autieri\orcidB}
\affiliation{International Research Centre Magtop, Institute of Physics, Polish Academy of Sciences,
Aleja Lotnik\'ow 32/46, PL-02668 Warsaw, Poland}

\date{\today}
\begin{abstract} 
We investigate the emergence of quantum anomalous Hall conductivity in a two-dimensional $d$-wave altermagnet on a Lieb lattice under an external magnetic field. Altermagnetic order induces momentum-dependent spin splitting without net magnetization in the relativistic limit, producing distinct spin-resolved bands at the $X$ and $Y$ valleys. The phase diagram features a normal insulator and a spin Chern insulator separated by an accidental Dirac semimetal. The magnetic field breaks rotational symmetry between valleys while maintaining vanishing total magnetization, enabling independent valley contributions to topology. One valley supports Chern numbers $C=-1$ or $0$, while the other hosts $C=0$ or $+1$, governed by field strength and bandwidth. This competition yields valley-dependent topology. Berry curvature analysis reveals fully gapped phases with total Chern numbers $C=\pm1$, separated by valley-selective gap closings. We uncover a mechanism for rapid magnetic control of the quantum anomalous Hall effect near the semimetal phase and highlight key distinctions from ferro-valleytronic and quantum spin Hall systems.
\end{abstract}


\maketitle
	
\section{Introduction}

The discovery of altermagnets has recently expanded the classification of magnetic materials,
offering a promising platform for spintronics. One of the most important characteristics of the altermagnets is the spin-momentum locking\cite{Smejkal2022PRX,doi:10.1021/acs.jpclett.5c03677,doi:10.1021/acs.jpclett.5c02854,doi:10.1021/acs.jpclett.5c03292,Tenzin2025,g32j-hnvz}, which, in the case of materials with valleys, is also named spin-valley locking with valleys that are connected by C$_n$ rotational symmetries. Therefore, it is also named C$_n$-paired spin-valley locking\cite{PhysRevX.15.021083,61s6-cbvn,Ma2024,zhang2025crystal}. The spin-valley in altermagnets has a non-relativistic origin, while there is a spin-valley from a relativistic origin, which was widely investigated in the last decades in 2D materials\cite{HUSSAIN2022169897} 
In particular, altermagnets can be divided into weak ferromagnets and pure altermagnets. Weak ferromagnets exhibit vanishing magnetization in the relativistic limit, while pure altermagnets are characterized by symmetry-protected spin textures that arise without net magnetic moments, also when spin–orbit coupling is included.

Recently, the field of 2D altermagnets attracted significant attention\cite{ma2021multifunctional,zou2025floquet,Gonzalez2025,Khan2025,SINGH2025101017}. 
One of the most investigated two-dimensional altermagnets is the Lieb lattice\cite{PhysRevLett.134.096703,durrnagel2025altermagnetic,wang2025valley,Xu2025,bezzerga2024giant,Khan2025} due to their abundant number of possible material realization\cite{10.1093/nsr/nwaf528}. While the majority of altermagnets exhibit weak ferromagnetism\cite{autieri2312staggered}, with the Néel vector aligned along the z-axis, the Lieb lattice realizes a pure altermagnet. With 
N\'eel vector in the ab-plane, the weak ferromagnetic moment appears in the ab-plane as well. Consequently, the anomalous Hall conductivity is always zero in the pristine Lieb lattice.
The topological properties of the Lieb lattice were recently investigated\cite{9wcm-pmr2,Ma2024}. It was proposed to be characterized by the mirror Chern number with analogy to the non-magnetic mirror Chern insulator and to host helical edge state as in the quantum spin Hall (QSH) phase\cite{PhysRevLett.134.096703,radhakrishnan2026topologicalpiezomagneticeffecttwodimensional,doi:10.1021/acs.jpclett.4c01724,doi:10.1021/acs.jpclett.2c03270}. On the other hand, the spin Chern number was proposed as a robust topological invariant in the material class of altermagnets\cite{s57q-q7gt,PhysRevB.111.224406,PhysRevB.111.085127}. The topological Lieb lattice exhibits Dirac-like surface states that are not at the $\Gamma$ point\cite{Sattigeri2025} even if what matters is the band inversion at the valley, like in the mirror Chern insulators. The surface Dirac points lie within bands of the same spin channels\cite{Deng2020}.

One of the most intriguing phenomena in magnetic topological systems is the quantum anomalous Hall effect (QAHE), which manifests as a quantization of the anomalous Hall conductivity in two-dimensional systems. 
This QAHE is well established in ferromagnetic materials near the topological band inversion, on both the topological and trivial sides of the topological transition.\cite{https://doi.org/10.1002/smll.201904322,Chang2016,Wang2015,PhysRevB.108.035121,PhysRevB.110.165112,majewski2026valleystopologicalphases}. 
Previous proposals for QAHE in systems with vanishing magnetization have largely relied on engineering rotational-symmetry breaking in altermagnets via strain\cite{Guo2023,10.1063/5.0147450,PhysRevB.107.214419} or by engineering the stacking\cite{Li2025} or surface\cite{doi:10.1021/acs.nanolett.5c05341}, to obtain a ferrimagnet. Interestingly, the same procedure also works if we start from an antiferromagnet with PT symmetry; also, in this case, we can obtain a ferrimagnetic system with a Chern number\cite{PhysRevResearch.2.022025}.
Such perturbations break the rotational symmetry and effectively convert the antiferromagnetic or altermagnetic state into a ferrimagnetic configuration, thereby enabling a finite Chern number with an induced net magnetization. 
Recently, it has been proposed that bilayers with broken PT symmetry can suppress non-relativistic spin-splitting through a combination of mirror and rotational symmetries. Therefore, this class of bilayers constitutes a system that lacks non-relativistic spin-splitting while exhibiting broken time-reversal symmetry. When spin–orbit coupling is included, the band degeneracy is lifted, making the system capable of hosting a Chern insulating phase.\cite{doi:10.1021/acs.nanolett.6c00290,doi:10.1021/acs.nanolett.3c02489,rn1l-d6cq}
Topological altermagnet can also exhibit layer Hall effect\cite{qin2026layerhalleffectinduced} and different kinds of high-order topology. One is an axion insulating phase; indeed, EuIn$_2$As$_2$ was shown to host g-wave magnetism in its collinear phase\cite{PhysRevB.108.075150,6yv6-kf97} and this g-wave magnetism allows different properties on different surfaces necessary to have chiral hinge states\cite{PhysRevLett.122.256402}. 
Additionally, higher-order topology has been studied as well, leading to corner states~\cite{9wcm-pmr2,kplp-819f}.

In this work, we demonstrate a route to realizing QAHE in a pure altermagnet via valleytronics. Instead of invoking structural distortions or symmetry-lowering strain, we apply an external magnetic field that preserves the intrinsic altermagnetic order while selectively breaking the fourfold rotational symmetry between inequivalent valleys. This magnetic-field-induced valley asymmetry allows each valley to contribute independently to the topological response, leading to valley-dependent Chern numbers and a net quantized Hall conductivity despite a vanishing total magnetization. 

\section{Model and symmetry analysis}

In this section, we set up the model Hamiltonian and discuss the properties of the valleytonics for this system with the allowed values of the Chern numbers.

\subsection{Model Hamiltonian}

We consider a Lieb lattice composed of two magnetic sublattices with antiparallel spin orientations along the z-axis. The Hamiltonian is formulated in the tensor-product space of sublattice ($\tau$) and spin ($\sigma$) degrees of freedom as:
\begin{align}
H(\mathbf{k}) =\;
& d_{0}(\mathbf{k}) \, \tau_{0} \otimes \sigma_{0}
+ d_{1}(\mathbf{k}) \, \tau_{x} \otimes \sigma_{0}
+ d_{3}(\mathbf{k}) \, \tau_{z} \otimes \sigma_{0}
\nonumber \\
&  
+ \Delta \, \tau_{z} \otimes \sigma_{z}
+ \lambda_{R}(\mathbf{k}) \, \tau_{y} \otimes \sigma_z,
\end{align}
where the first three terms are hopping parameters, $\Delta$ represents the on-site spin-splitting and $\lambda_R(\mathbf{k})$ is the Rashba term which breaks the inversion symmetry. The on-site spin-splitting $\Delta \tau_z \otimes \sigma_z$ enforces antiparallel spin polarization along the z-axis on the two sublattices. Once we add the $d_3\tau_z\sigma_0$ hopping, the interplay between the two breaks time‑reversal symmetry while preserving zero net magnetization in the non-relativistic limit.

The scalar term is:
\begin{equation}
d_{0}(\mathbf{k}) = \tfrac{1}{2}\big(\epsilon_{1}(\mathbf{k}) + \epsilon_{2}(\mathbf{k})\big)
\end{equation}
originates from the intra-sublattice hopping, $\epsilon_{\alpha}(\mathbf{k})=2t_{\alpha x}cos(k_x)+2t_{\alpha y} cos(k_y)$, and does not affect the band topology. The nearest-neighbor inter-sublattice hopping is encoded in
\begin{equation}
d_{1}(\mathbf{k}) = 4t \cos\!\left(\tfrac{k_x}{2}\right)\cos\!\left(\tfrac{k_y}{2}\right),
\label{Eq:2}
\end{equation}
The d$_3$ term is
\begin{equation}
d_{3}(\mathbf{k}) =
\tfrac{1}{2}\big(\epsilon_{1}(\mathbf{k}) - \epsilon_{2}(\mathbf{k})\big)
+ M_{0} + M_{1}(\cos k_x - \cos k_y)
\end{equation}
and owing to the hopping condition
\begin{equation}
t_{\alpha x} = -t_{\alpha y} = (-1)^\alpha t_d^\alpha,
\end{equation}
the difference $\epsilon_1(\mathbf{k})-\epsilon_2(\mathbf{k})$ transforms as a $d_{x^2-y^2}$ form factor and changes sign under $C_4$ rotation.
The two contributions, therefore, enter the Hamiltonian only through their sum,
\begin{equation}
d_{3}(\mathbf{k}) = M_{0} + M_{1}^{\mathrm{eff}}(\cos k_{x}-\cos k_{y}),
\qquad
M_{1}^{\mathrm{eff}} = M_{1} - 2 t_{d}.
\end{equation}
Therefore, the momentum‑dependent part of $d_3(\mathbf{k})$ encodes the d-wave altermagnetic structure and transforms according to the $d_{x^{2}-y^{2}}$ irreducible representation of the square-lattice point group. In what follows, we therefore treat $M_{1}^{\mathrm{eff}}$ as the relevant $d$-wave mass parameter controlling band inversion and the associated topological transitions. The spin-momentum locking d$_{x2-y2}$ in the non-relativistic limit allows for two valleys at X and Y with opposite spin-resolved band structures\cite{fan2025valley}. 
The term $M_0$ introduces a staggered on-site energy, effectively acting like a magnetic field that enables the QAHE. As a result, the system exhibits a ferrimagnetic behavior.
The Kane-Mele spin-orbit coupling is described as Kane-\cite{PhysRevLett.134.096703}
\begin{equation}\label{Eq_SOC}
H_{SOC}=\lambda\sin\tfrac{k_x}{2}\sin\tfrac{k_y}{2} \tau_y \sigma_z,
\end{equation}
among the high-symmetry points, this term is nonzero only at the M point. 
In what follows, we set t=1 and $\Delta=1$.

\begin{figure*}
    \centering
    \includegraphics[width=0.99\linewidth,height=0.7\linewidth]{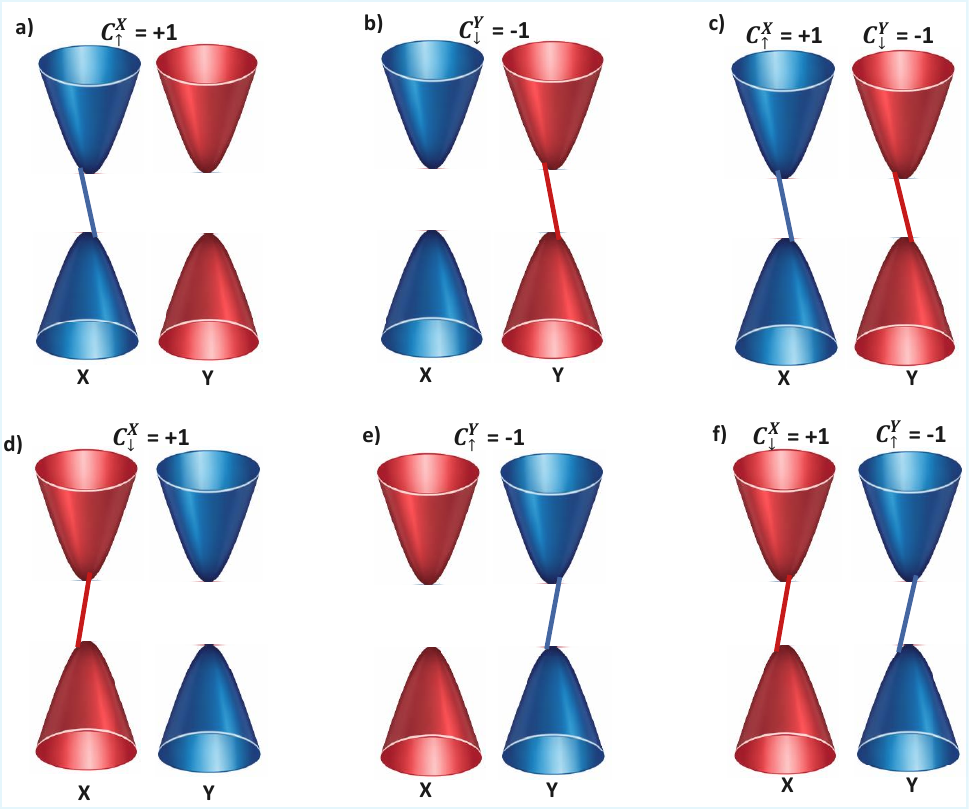}
    \caption{Schematic illustration of topological surface states connecting valence and conduction bands for different topological phases and their associated spin-resolved Chern numbers and valleys. Chern numbers not reported in the panels are zero. Blue and red paraboloids represent the valleys with spin-up and down, respectively. For $t_d > 0$, the possible phases are: (a) $C_{\uparrow}^{X} = +1$, (b) $C_{\downarrow}^{Y} = -1$, and (c) the combination of the previous two cases with $C_{\uparrow}^{X} = +1$ and $C_{\downarrow}^{Y} = -1$.  For $t_d < 0$, the possible phases are: (d) $C_{\downarrow}^{X} = +1$, (e) $C_{\uparrow}^{Y} = -1$, and (f) the combination of the previous two cases with $C_{\downarrow}^{X} = +1$ and $C_{\uparrow}^{Y} = -1$.}
    \label{Figure1} 
\end{figure*}

\subsection{Valleytronics and Chern number}

The quantum spin Hall insulator in non-magnetic systems is characterized by a $\mathbb{Z}_2$ topological invariant\cite{PhysRevLett.95.146802,https://doi.org/10.1002/aelm.202300156}, protected by time-reversal symmetry and with Dirac surface states at the $\Gamma$ point. In contrast, in the altermagnetic Lieb lattice, we describe this phase by the spin Chern number. The Chern number characterizes topological phases that are not protected by any symmetry.
From the Topological Kirchhoff law and bulk-edge correspondence\cite{PhysRevB.88.161406}, we can define different kinds of Chern numbers as reported in the supplementary Materials. Due to the combined symmetry C$_{4z}$T of the investigated altermagnetic system for zero magnetic field, the spin reverses at the valleys; therefore, we have opposite spins at the two valleys from which we have:
\begin{equation}\label{constraints}
    C_{\downarrow}^X=-C_{\uparrow}^Y \quad C_{\downarrow}^Y=-C_{\uparrow}^X
\end{equation}
where $X=(\pi,0)$, and $Y=(0,\pi)$.
These relationships force the total Chern number C to be zero; however, the system hosts valleys with opposite Chern numbers, enabling potential valleytronic applications.
Therefore, at zero magnetic field, C=C$_{sv}$=0 while C$_s$=C$_v$, where C$_v$ accidentally coincides with the mirror Chern number for this class of materials. Moreover, at most one member of each spin--valley pair can become topologically active. When this occurs, the system realizes a spin Chern insulator with C$_s$=2.  Upon applying a finite magnetic field $M_0$, the constraints in equations(\ref{constraints}) no longer hold, and the system can transition into a Chern insulator phase with C$\neq$0.

In one region of the phase diagram, we have that C$_{\downarrow}^X$=C$_{\uparrow}^Y$=0, there are only two independent topological invariants in the model, which are C$_{\uparrow}^X$=-1,0 and C$_{\downarrow}^Y$=1,0. This gives rise to four different topological phases. The first is the trivial phase, where C$_{\uparrow}^X$=C$_{\downarrow}^Y$=0. There are two Chern insulating phases for C=C$_{\uparrow}^X$=-1 or for C=C$_{\downarrow}^Y$=1, in which both of them C$_{sv}$=-1. The fourth is characterized by C$_{\uparrow}^X$=-1 and C$_{\downarrow}^Y$=1 from which we get C=C$_{sv}$=0 and $|C_{s}|$=$|C_{s}|$=2. From the last relation, we understand that in our model, the spin Chern insulator is equivalent to a valley Chern insulator in this material class.  In our phase diagram, we will describe these four phases by C=0, C=1, C=-1 and $|C_{s}|$=2.
The band structure of these three possible topological phases is shown in Fig.~\ref{Figure1}(a--c). Figure~\ref{Figure1}(a) corresponds to a phase in which the nonzero Chern number appears in the spin-up channel, while Fig.~\ref{Figure1}(b) corresponds to a phase where the nonzero Chern number is in the spin-down channel. Figure~\ref{Figure1}(c) represents the phase in which topological surface states are present in both spin channels. When the magnetic field $M_0$ is zero, this latter phase is the only symmetry-allowed topological phase.
In another region of the phase diagram, the spins are not reversed; instead, the band structure is inverted. In this case, the three topological phases shown in Fig.~\ref{Figure1}(d--f) arise.
The number of topological surface states connecting the valence and conduction bands is equal to $|C_s|$ in all cases, from which we can confirm $|C_s|$ as a robust topological invariant for this material class.

\begin{figure}
    \centering
    \includegraphics[width=0.99\linewidth]{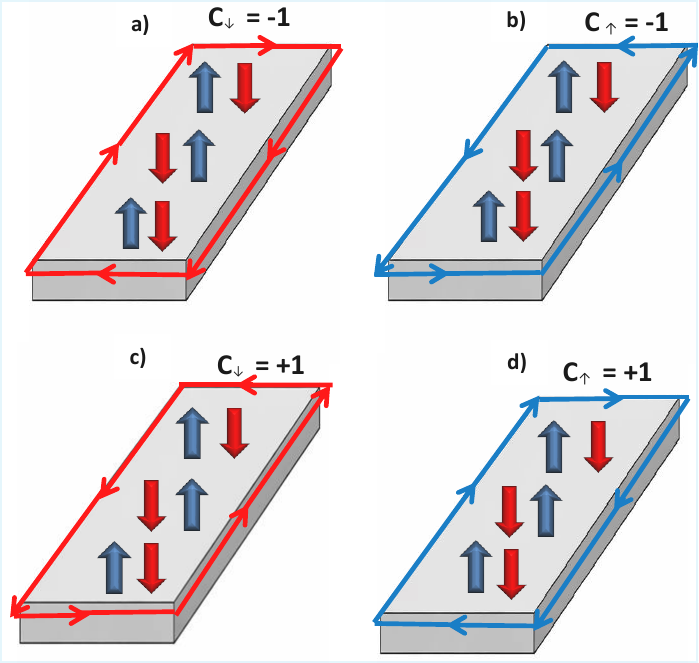}
    \caption{Schematic of edge states in the real space of a monolayer QAH insulator with Chern number $C = C_{\uparrow} + C_{\downarrow}$ for different cases: (a) $C_{\downarrow} = +1$, (b) $C_{\uparrow} = +1$, (c) $C_{\downarrow} = -1$, and (d) $C_{\uparrow} = -1$. Red and blue arrows on the gray sample indicate the magnetic atoms. The arrows on the edges indicate the movement of the spinful electrons. For the case with C$_{s}$=2, we need to combine two of them.}
    \label{Figure2} 
\end{figure}

We report the edge states in real space for the Chern insulator in Figure \ref{Figure2}.
The possible four configurations of the chiral edge states within this model are reported in Fig. \ref{Figure1}(a-d). At the same time, the non-trivial band structure with the topological surface states connecting conduction and valence band can have four possible configurations reported in \ref{Figure2}(a-d). In the case of the Chern valley insulator, we have both spin-up and spin-down with opposite Chern numbers, with the system behaving like a QSH insulator\cite{PhysRevLett.134.096703}.

In ferrovalley materials\cite{Li2024}, $C^X = C^Y$; therefore, the valley Chern number is zero, and the spin Chern number is equal to the Chern number if one of the two spin channels has a trivial Chern number, as usually happens. Consequently, the Chern number alone is sufficient to describe all possible configurations. Since strain-induced and correlation-driven topological transitions have been observed in ferrovalley materials\cite{Hussain2023,PhysRevB.109.075147}, we expect similar behavior to occur in altermagnetic valleytronics.

\section{Results}

\subsection{Analytical Results}

The topological properties of the altermagnetic model originate from the
structure of the spin-resolved Dirac masses along the Brillouin zone boundary. In the absence of spin–orbit coupling (SOC), the Hamiltonian
decouples into two spin sectors $s=\pm1$,
\begin{equation}
H_{s}(\mathbf{k}) = d_{1}(\mathbf{k})\,\tau_{x}
+ m_{s}(\mathbf{k})\,\tau_{z},
\end{equation}
where the inter-sublattice hopping, Eq.~\ref{Eq:2} vanishes whenever $k_{x}=\pi$ or $k_{y}=\pi$. Thus, the entire $X$--$M$--$Y$ boundary forms a nodal line of the kinetic term, where M=($\pi$,$\pi$). Along this boundary, the spectrum is governed solely by the Dirac mass
\begin{equation}
m_{s}(\mathbf{k}) = M_{0} + M_{1}^{\mathrm{eff}}(\cos k_{x}-\cos k_{y}) + s\Delta.
\end{equation}
Band touchings occur wherever $m_{s}(\mathbf{k})=0$. Because
$\cos k_{x}-\cos k_{y}$ varies continuously along the boundary, the
solutions of $m_{s}(\mathbf{k})=0$ are, in general, located at
parameter-dependent momenta. Along the $X$--$M$ line , the
condition becomes, see supplementary information for further data,
\begin{equation}
\cos k_{y}^{\ast}
= -1 + \frac{M_{0}+s\Delta}{M_{1}^{\mathrm{eff}}},
\end{equation}
while along the $Y$--$M$ line it becomes
\begin{equation}
\cos k_{x}^{\ast}
= -1 - \frac{M_{0}+s\Delta}{M_{1}^{\mathrm{eff}}}.
\end{equation}
Whenever the right-hand sides lie within $[-1,1]$, the corresponding spin
sector hosts a Dirac point at $(\pi,k_{y}^{\ast})$ or $(k_{x}^{\ast},\pi)$.
Thus, except in special limits, the Dirac cones are not pinned at the
high-symmetry points $X$ or $Y$, but instead move continuously along the
zone boundary as $M_{0}$ or $M_{1}^{\mathrm{eff}}$ is tuned. For
fine-tuned values of the parameters, the entire boundary can satisfy
$m_{s}(\mathbf{k})=0$, producing an accidental Dirac nodal line. This accidental nodal line divides the two robust phases of the altermagnetic normal insulator and altermagnetic spin Chern insulator. 

SOC qualitatively modifies this picture. The Rashba term vanishes exactly at $X$ and $Y$ but is finite everywhere else along the
boundary. As a result, the SOC gaps the shifted Dirac points at
$(\pi,k_{y}^{\ast})$ and $(k_{x}^{\ast},\pi)$, generating sharply localized
Berry-curvature peaks at their actual positions. Only at $X$ and $Y$ does
SOC vanish identically, so these points remain strictly mass-controlled.
Consequently, in the presence of SOC, the topological transitions are
governed by the sign of the mass at the true Dirac points rather
than by the masses at $X$ and $Y$ alone.

Each gapped Dirac cone contributes $\pm\tfrac12$ to the Chern number,
with the sign determined by the chirality and by the sign of the mass at
the corresponding band-touching momentum. Because the Dirac points move
along the boundary, the Chern number changes when $m_{s}(\mathbf{k}^{\ast})$
changes sign. The simplified expression
\begin{equation}
C = \frac{1}{2}\left[\operatorname{sgn}m(Y)-\operatorname{sgn}m(X)\right]
\end{equation}
is valid only when the Dirac cones sit exactly at $X$ and $Y$; in the generic case, the correct topological index must be evaluated at the
shifted Dirac points.

Because these momenta depend on $M_{0}$ and
$M_{1}^{\mathrm{eff}}$, the Berry-curvature hotspot at $\mathbf{k}$=$\mathbf{k}^{\ast}$ moves along the
$X$--$M$--$Y$ boundary as the parameters are tuned. This motion explains why the Berry curvature dipole cannot be approximated by
$\Omega(X)-\Omega(Y)$: its magnitude and sign depend sensitively on the
actual location of the Berry-curvature pole and on how the Fermi
surface intersects this moving hotspot.

A particularly transparent limit arises when $M_{1}^{\mathrm{eff}}=0$, in
which case the Dirac cones are pinned at $X$ and $Y$ with masses
$m_{X}=M_{0}-\Delta$ and $m_{Y}=M_{0}+\Delta$. The condition for a
topological phase reduces to $|M_{0}|<\Delta$, corresponding to a
``pure altermagnetic QAHE'' driven solely by the uniform exchange field.
Conversely, when $M_{1}^{\mathrm{eff}}\neq0$, the Dirac cones shift away
from $X$ and $Y$, SOC gaps them at their displaced positions, and the
topology is controlled by the sign structure of $m_{s}(\mathbf{k}^{\ast})$.

\begin{figure}
    \centering
    \includegraphics[width=0.99\linewidth]{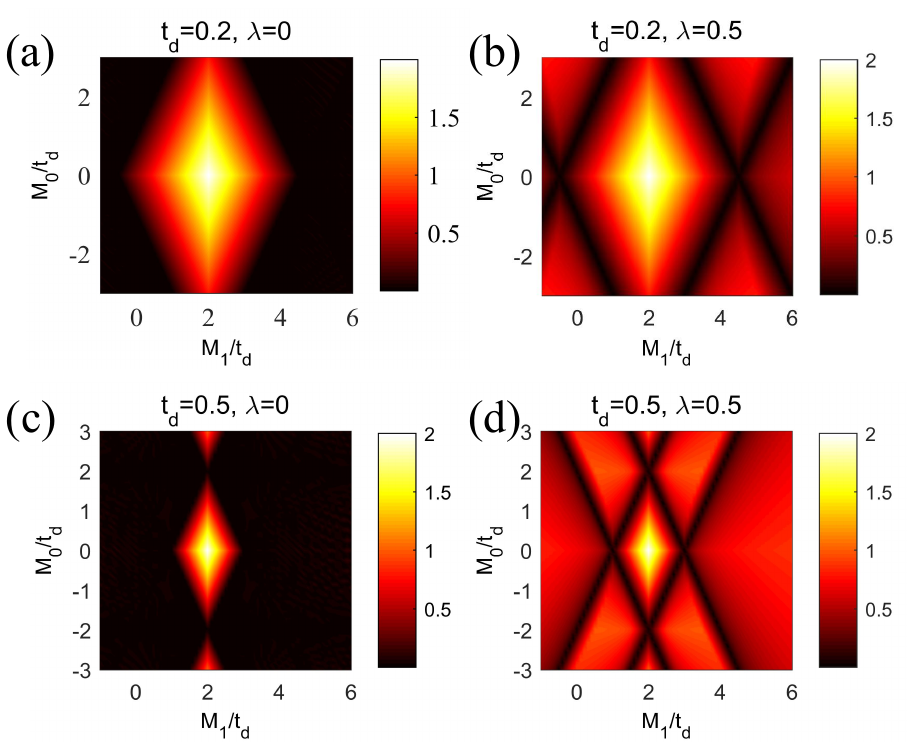}
    \caption{Global band gap $E_{\mathrm{gap}}$ as a function of the masses $(M_{0},M_{1})$ for two representative values of the altermagnetic hopping anisotropy $t_{d}$. (a),(b) $t_{d}=0.2<\Delta/4$ without and with SOC, respectively. The gap closes only along straight lines corresponding to Dirac mass inversions at the $M$, $X$, and $Y$ points in the absence of SOC, while SOC removes the $M$-point transition and confines all gap closings to $X$ and $Y$. (c),(d) $t_{d}=0.5>\Delta/4$ without and with SOC, respectively. For $t_{d}>\Delta/4$ a gapless region appears around the origin due to spin-polarized Weyl points on the Brillouin-zone edges; SOC gaps these Weyl points and collapses the gapless region onto the mass-inversion lines $m(X)=0$ and $m(Y)=0$, in agreement with the analytical theory.}
    \label{fig:gapmap} 
\end{figure}

\subsection{Numerical Results}

We now corroborate the analytical picture by performing numerical calculations
of the band structure, global band gap, Berry curvature, and Chern number on a
discrete Brillouin-zone mesh. The numerical results confirm the key analytical
finding that the Dirac points of the altermagnetic model are not fixed at the
high-symmetry points $X$ or $Y$, but instead move continuously along the
$X$--$M$--$Y$ boundary as a function of the masses $(M_{0},M_{1})$ and
the altermagnetic anisotropy $t_{d}$. This motion plays a central role in
determining the gap structure and the topological phase diagram.

Figure~\ref{fig:gapmap} shows the global band gap $E_{\mathrm{gap}}$ as a
function of $(M_{0},M_{1})$ for two representative values of $t_{d}$, both
without and with SOC. For $t_{d}=0.2<\Delta/4$
[Figs.~\ref{fig:gapmap}(a),(b)], the system is insulating over a wide region of
parameter space. In the absence of SOC, the gap closes only along narrow lines
that coincide with the analytical conditions.  When SOC is included, the gapless
regions collapse onto the momenta where the SOC form factor vanishes, namely
$X$ and $Y$, in agreement with the analytical result.

For $t_{d}=0.5>\Delta/4$ [Figs.~\ref{fig:gapmap}(c),(d)], the gap map changes
qualitatively. Without SOC, a finite region around the origin becomes gapless,
reflecting the emergence of mobile Dirac points that sweep along the
$X$--$M$--$Y$ boundary as $(M_{0},M_{1})$ are varied. In this regime, the
Dirac-point conditions $m_{s}(\mathbf{k}^{\ast})=0$ are satisfied over extended
segments of the boundary, producing accidental nodal lines in momentum space.
Only sufficiently large $|M_{0}|$ or $|M_{1}|$ shifts the Dirac masses away from
zero and reopen a global gap. When SOC is added, these nodal lines are gapped
everywhere except at $X$ and $Y$. As a result, the gapless region collapses onto the mass-inversion lines $m(X)=0$ and $m(Y)=0$. Further insight is provided by the band structures shown in the Supplementary
Information. Figure~S1 displays the evolution of the spectrum as a function of $t_{d}$,
both without and with SOC and with vanishing masses. For
$t_{d}=0.2<\Delta/4$, the bands are fully gapped. At the critical value
$t_{d}=0.25\Delta$, the Dirac points sit exactly at $X$ and $Y$, and SOC cannot
open a gap. For
$t_{d}=0.5>\Delta/4$, the Dirac points migrate away from $X$ and $Y$ and appear
along the $X$--$M$ and $M$--$Y$ segments, forming accidental nodal lines along
the zone boundary. SOC gaps these crossings everywhere except at $X$ and $Y$.

\begin{figure}[t]
    \centering
    \includegraphics[width=0.99\linewidth]{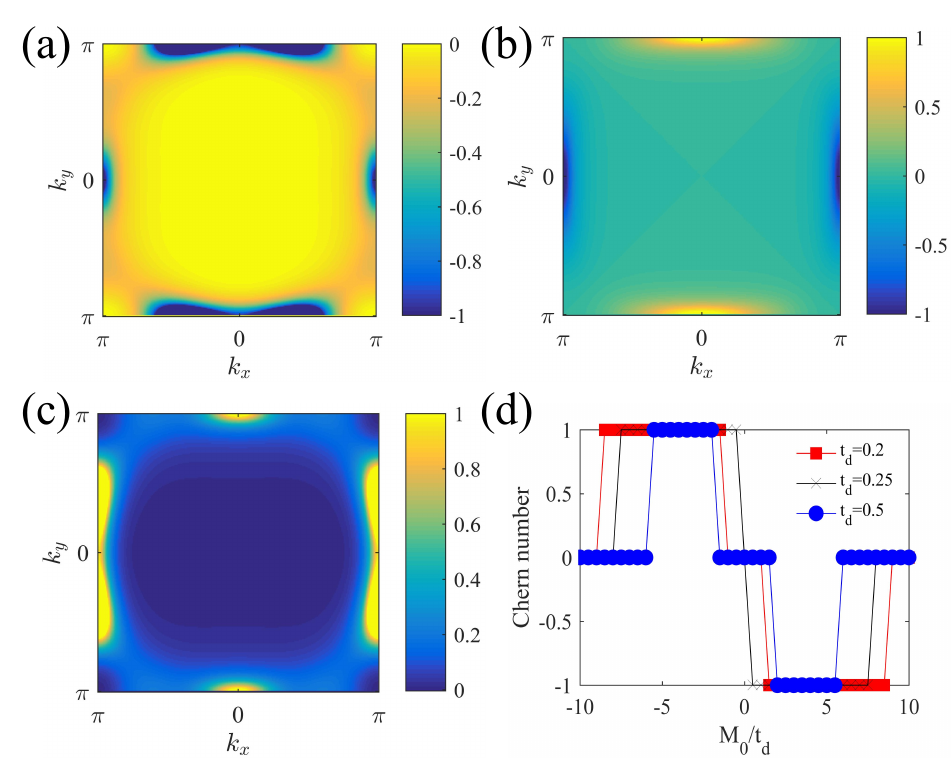}
    \caption{
     (a)–(c) Berry curvature $\Omega(\mathbf{k})$ of the occupied bands for 
    $t_{d}=0.5$, $\lambda=0.5$, $M_{0}=-t_{d}$, and 
    $M_{1}=t_{d},\,2t_{d},\,3t_{d}$, respectively. 
    These three parameter sets realize Chern numbers $C=-1,0,+1$, as shown in 
    Fig.~S2. For $M_{1}=t_{d}$, the curvature is dominated by negative hot spots 
    near the $Y$ valley, yielding a net negative flux. At $M_{1}=2t_{d}$ the 
    valley contributions nearly cancel and the integrated curvature vanishes, 
    while for $M_{1}=3t_{d}$ the hot spots reverse sign and produce a positive 
    net flux. (d) Chern number $C$ as a function of $M_{0}$ for $t_{d}=0.5$, 
    $\lambda=0.5$, and $M_{1}=0$. Quantized plateaus at $C=\pm1$ are separated 
    by a region with $C=0$.}
    \label{fig:berry} 
\end{figure}

Figure~S2 focuses on the case $t_{d}=0.5$, $M_{0}=-t_{d}$ and illustrates how
the Chern number is controlled by the sign of the Dirac mass at the
true Dirac points.  For $M_{1}=t_{d}$, the Dirac point lies along
the $M$--$Y$ edge and is gapped by SOC, yielding a Chern number $C=-1$. For
$M_{1}=2t_{d}$, the Dirac masses at the relevant momenta share the same sign,
the spectrum is fully gapped, and the phase is topologically trivial with
$C=0$. For $M_{1}=3t_{d}$, the Dirac point migrates to the $M$--$X$ edge, and
the SOC gaps it into a phase with $C=+1$. These three cases provide a direct
numerical realization of the analytical rule that the Chern number is
determined by the sign structure of the Dirac masses at the parameter-dependent
Dirac points. A closer inspection of the valley-resolved Berry curvature reveals an additional layer of structure that is fully consistent with the
Dirac-mass topology. For $M_{0}<0$, the nontrivial Chern number is
generated exclusively by the spin-up sector, whereas for $M_{0}>0$ the spin-down sector becomes topologically active. In both cases,
only a single spin channel contributes to the Berry curvature, and
the opposite spin sector remains topologically trivial. This spin
selectivity originates from the fact that the sign of $M_{0}$ enters
the Dirac mass $m_{s}(\mathbf{k})$ with opposite weights for the two
spin sectors, thereby determining which spin species undergoes the
valley mass inversion responsible for the QAHE.

Figure~S3 provides a direct visualization of the analytical Dirac-point
conditions by mapping the momentum positions $k_{x}^{\ast}$ and $k_{y}^{\ast}$
for which band touching occurs in the absence of SOC. The plots show, for each
spin sector and for both the $X$--$M$ and $Y$--$M$ directions, the regions in
the $(M_{0},M_{\mathrm{eff}})$ plane where the equation
$m_{s}(\mathbf{k}^{\ast})=0$ admits a real solution. For weak altermagnetic
anisotropy, [Fig.~S3(a)], the allowed regions are narrow and
strongly spin-selective: the spin-up and spin-down Dirac points appear in
disjoint portions of parameter space and never occur simultaneously near
the same valley. As a result, the Brillouin-zone boundary is never close to a
spin-degenerate band touching, and accidental nodal-line behavior does not
emerge. In contrast, for stronger anisotropy, $t_{d}=0.5\Delta$ [Fig.~S3(b)], the
structure changes qualitatively. The allowed regions expand dramatically, and
for sufficiently large $|M_{0}|$ or $|M_{\mathrm{eff}}|$, the spin-up and
spin-down Dirac points appear in the same region of the $X$--$M$ or
$Y$--$M$ boundary. This overlap indicates that both spin sectors satisfy the
mass-inversion condition at nearby momenta, signaling proximity to an accidental
Dirac nodal-line regime.

Figure~S4 illustrates the emergence of an accidental Dirac nodal line (DNL) in
the fine-tuned limit where $M_{1}^{\mathrm{eff}}=0$ and $M_{0}=-S_{0}\Delta$.
In the absence of SOC, the inter-sublattice hopping $d_{1}(\mathbf{k})$
vanishes along the entire Brillouin-zone boundary, so the spectrum on this
boundary is controlled solely by the Dirac mass $m_{s}(\mathbf{k})$. For the
chosen parameters, the mass of the spin sector $s=S_{0}$ satisfies
$m_{S_{0}}(\mathbf{k})=0$ along the entire $X$--$M$--$Y$ path, producing a
continuous band touching and a Dirac nodal line. When SOC is introduced, the
nodal line is gapped everywhere except at $X$ and $Y$, where the SOC form
factor vanishes. These two symmetry-protected Dirac points are precisely the
momenta that control the topological phase transitions in the SOC-gapped
system, in agreement with the analytical condition that the Chern number is
determined by the sign structure of the valley Dirac masses.

Figure~\ref{fig:berry} illustrates the momentum-space structure of the Berry
curvature and its direct connection to the Chern number. Panels (a)–(c) show
the Berry curvature $\Omega(\mathbf{k})$ for $t_{d}=0.5$, $\lambda=0.5$,
$M_{0}=-t_{d}$, and $M_{1}=t_{d}, 2t_{d}, 3t_{d}$. These parameters correspond to
the band structures of Fig.~S2 and realize Chern numbers $C=-1,0,+1$,
respectively. For $M_{1}=t_{d}$, the curvature is concentrated near the
parameter-dependent Dirac point along the $M$--$Y$ edge, yielding a net
negative flux. At $M_{1}=2t_{d}$ the valley contributions nearly cancel,
yielding $C=0$. For $M_{1}=3t_{d}$, the curvature hot spots reverse sign,
producing a positive net flux and $C=+1$. These textures provide a direct
momentum-space visualization of the Dirac-mass topology and the
valley-resolved Berry-curvature dipole underlying the QAHE.

The global evolution of the Chern number is summarized in
Fig.~\ref{fig:berry}(d), which shows $C$ as a function of $M_{0}$ for fixed
 $\lambda=0.5$, $M_{1}=0$ and different $t_d$. The Chern number exhibits quantized
plateaus at $C=\pm1$ separated by a region with $C=0$, and the
transitions occur precisely at the values of $M_{0}$ where the Dirac masses at
the relevant Dirac points change sign. Results show that an increase in hopping results in an increase in the threshold $M_0$ where the Chern number is nonzero. This behavior corresponds with the analytical condition that the QAHE is controlled by the
sign structure of the Dirac masses at the parameter-dependent Dirac points.

The momentum-space distribution of the Berry curvature further
reflects the valley mechanism. For $C=-1$, the curvature is dominated by a
negative hotspot centered near the $Y$ valley, while for $C=+1$ the
hotspot shifts to the vicinity of the $X$ valley with opposite sign.
Importantly, these hotspots are not sharply localized at the
high-symmetry points themselves. Instead, they appear as broadened
features spread around the valleys. This behavior is a direct
consequence of the fact that, in the topological regime, the band
touching that triggers the mass inversion does not occur exactly at
$X$ or $Y$, but rather at nearby momenta along the $X$--$M$ or
$Y$--$M$ directions where the SOC form factor is finite.  The resulting valley-centered but spatially extended curvature profile provides a direct numerical signature of the spin-selective Dirac mass inversion that underlies the nonzero Chern
number.

\begin{figure*}[t]
    \centering
    \includegraphics[width=0.99\linewidth]{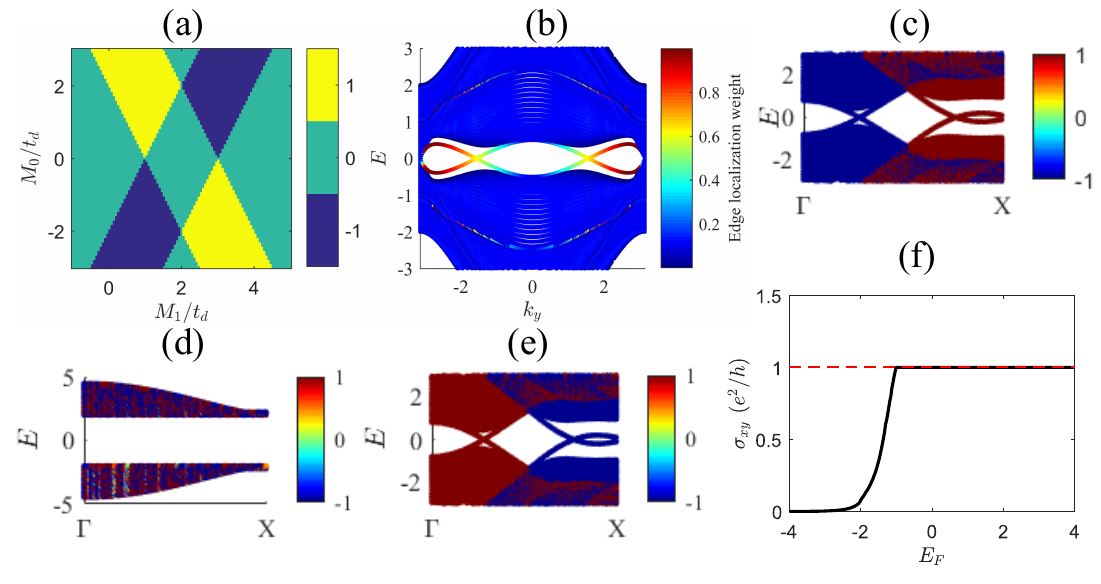}
    \caption{
      (a) Chern-number phase diagram $C(M_{0}/t_{d},M_{1}/t_{d})$ for $t_{d}=0.5$ and $\lambda_{R}=0.5$. 
    (b) Ribbon spectrum for a strip open along $x$ and periodic along $y$ at 
    $(M_{0},M_{1})=(-t_{d},3t_{d})$, corresponding to a point inside the 
    $C=1$ region. The color scale denotes edge-localization weight and reveals 
    a single chiral edge mode per boundary. 
    Edge states for region with $C=0$, but (c) and (e) nonzero $C_s$ and (d) $C_s=0$.  Color bar denotes $S_z$.
    (f) Hall conductivity $\sigma_{xy}(E_{F})$ in units of $e^{2}/h$ for 
    $(M_{0},M_{1})=(-t_{d},3t_{d})$. A quantized plateau at $\sigma_{xy}=1$ 
    appears when the Fermi energy lies inside the bulk gap, consistent with the 
    Chern number and the edge-state structure.}
    \label{fig:chern} 
\end{figure*}

To establish the topological character of the altermagnetic phase, we combine
bulk, boundary, and transport diagnostics in Fig.~\ref{fig:chern}. The
Chern-number map in Fig.~\ref{fig:chern}(a), obtained from the
Fukui--Hatsugai--Suzuki algorithm \cite{fukui2005chern}, reveals a well-defined region in the
$(M_{0}/t_{d},M_{1}/t_{d})$ plane where the system acquires a nontrivial Chern
number $C=1$. This topological lobe emerges from the interplay between the
$d$-wave altermagnetic mass term and Rashba SOC, and it is separated from the
trivial $C=0$ regime by sharp phase boundaries.  An important feature of the phase diagram in Fig.~\ref{fig:chern}(a) is that
the Chern number is not symmetric under $M_{0}\!\to\!-M_{0}$. For example, at
$(M_{0},M_{1})=(-t_{d},t_{d})$ we obtain $C=-1$, whereas at
$(M_{0},M_{1})=(+t_{d},t_{d})$ the Chern number changes sign to $C=+1$. This
behavior reflects the fact that $M_{0}$ enters the Hamiltonian in the same mass
channel as the altermagnetic term $M_{1}(\cos k_{x}-\cos k_{y})$. Consequently,
changing the sign of $M_{0}$ reverses the sign of the momentum-dependent mass
responsible for the band inversion, and therefore reverses the sign of the
Berry curvature throughout the Brillouin zone. This mechanism is analogous to
the mass-sign reversal in the Haldane \cite{haldane1988model} and Qi--Wu--Zhang \cite{qi2006topological} models, where the Chern
number changes sign when the Dirac mass is inverted.

The boundary spectra in Fig.~\ref{fig:chern}(b) provide direct evidence for the bulk--boundary correspondence. For
$(M_{0},M_{1})=(-2t_{d},3t_{d})$, a point deep inside the $C=1$ region, the
ribbon spectrum exhibits a single chiral edge mode per boundary traversing the bulk gap wih the formation of Dirac points between bands of the same spin channel. These two Dirac points symmetric with respect to the $\Gamma$ point were also reported from ribbons calculated ab initio\cite{Sattigeri2025}. 
Our results reveal that even in the absence of $M_0$, $M_1$ can control topology, especially spin Chern topology. Figs.~(c)-(e) demonstrate the edge state for regions with Chern number of zero in Fig.~\ref{fig:chern}(a), but, there is a spin Chern number for $M_1=-t_d$ (Fig.~\ref{fig:chern}c) and $M_1=4t_d$ (\ref{fig:chern}(e)), while the spin chern number is zero for $M_1=2t_d$ as we can see in Fig.~\ref{fig:chern}(d). These results are consistent with the recent literature~\cite{PhysRevLett.134.096703}.  

The transport response shown in Fig.~\ref{fig:chern}(f) completes the picture.
For the same parameters as in Fig.~3(b), the Hall conductivity
$\sigma_{xy}(E_{F})$ exhibits a quantized plateau at $\sigma_{xy}=e^{2}/h$
whenever the Fermi energy lies inside the bulk gap. The plateau collapses only when $E_{F}$ enters the bulk bands, consistent with the expected behavior of a Chern-insulating state. The agreement between the Chern number, the chiral edge spectrum, and the quantized Hall response demonstrates that the altermagnetic system realizes a robust quantum anomalous Hall phase.

The conductivity map in Fig.~S5 clearly shows that a wide
parameter window in the $(M_{0},\mu)$ plane supports a nearly quantized
Hall response with $\sigma_{xy}\!\approx\! e^{2}/h$ at fixed $M_{1}=3t_{d}$.
This plateau coincides with the region where the system lies deep inside
the $C=1$ insulating phase, confirming that the quantized anomalous Hall
effect remains robust against moderate variations of both the mass term
$M_{0}$ and the Fermi level. The rapid collapse of the plateau when $\mu$
enters the bulk bands or when $M_{0}$ approaches the topological phase
boundary directly visualizes the stability range of the QAH phase and
highlights the energetic protection of the chiral edge channel.

\begin{figure}[t]
    \centering
    \includegraphics[width=0.99\linewidth]{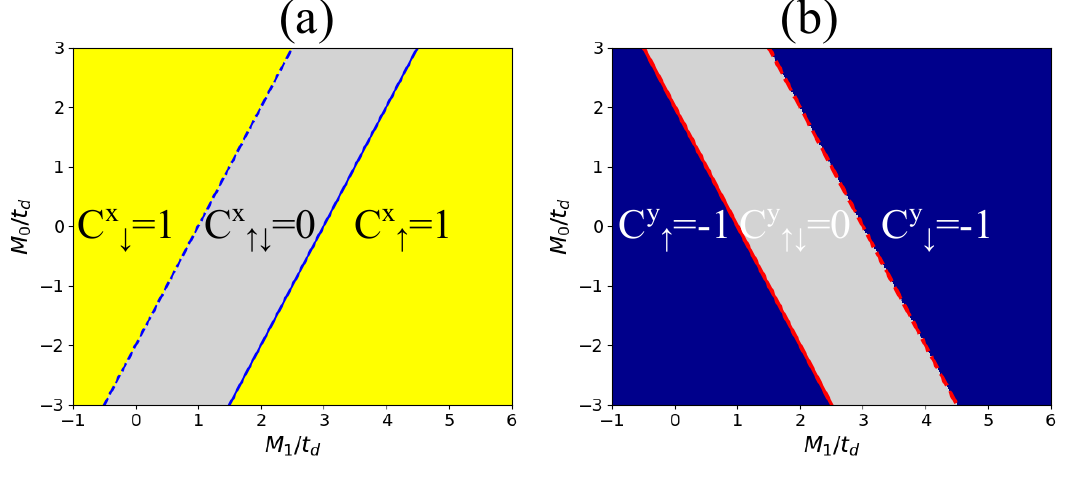}
    \caption{
 Phase boundaries associated with the gap-closing conditions in the 
(a) $X$ valley and (b) $Y$ valley, plotted in the $(M_{1}/t_{d},\, M_{0}/t_{d})$ parameter plane. 
The solid and dashed lines correspond to the spin-up ($\sigma=+1$) and spin-down ($\sigma=-1$) branches of the valley-dependent gap-closing equations, respectively. 
The colored regions indicate the parameter domains where the corresponding spin–valley Chern numbers acquire nonzero values, while the gray region denotes the topologically trivial regime. 
These boundaries determine the onset of valley-resolved topological phases discussed in the main text.
}
\label{fig:valleygaps}
\end{figure}

A direct comparison between the global band gaps at the two
high–symmetry valleys, shown in Fig.~\ref{fig:valleygaps}(a,b) and
Fig.~S6, reveals a transparent correspondence between the
topological phase boundaries and the underlying valley physics of
the altermagnetic state. At the $X$ and $Y$ points, the Rashba SOC
form factor vanishes identically, so SOC cannot open a gap at
these momenta. Consequently, the local Dirac masses at the valleys
are governed entirely by the altermagnetic terms, and the
spin–resolved band–touching conditions reduce to the simple
analytic relations
\begin{align}
\frac{M_{0}-2M_{1}}{t_{d}} &= -\sigma\,\frac{\Delta}{t_{d}} - 4
\quad (X\text{ valley}), \\
\frac{M_{0}+2M_{1}}{t_{d}} &= 4 - \sigma\,\frac{\Delta}{t_{d}}
\quad (Y\text{ valley}),
\end{align}
with $\sigma=\pm 1$ denoting spin up/down. These equations define
straight zero–gap lines in the $(M_{0},M_{1})$ plane, separated by
$2\Delta/t_{d}$ for the two spin channels and centered at $-4$ and
$+4$ for the $X$ and $Y$ valleys, respectively.

Importantly, the topological phase boundaries in
Fig.~\ref{fig:valleygaps}(a,b) do \emph{not} coincide with these
spin–resolved valley gap closings. Because SOC is inactive exactly
at $X$ and $Y$, a band touching at the valley itself does not
change the Chern number. The nontrivial phases ($C=\pm 1$) emerge
only when the Dirac crossing is displaced away from the valley
along the $X$--$M$ or $Y$--$M$ directions, where SOC becomes
finite and gaps the band crossing. Thus, the topological transition is
triggered by a sign reversal of the altermagnetic mass at nearby
momenta where SOC is operative, rather than by a gap closing at
the high–symmetry points.

Overlaying the Chern–number regions on the valley gap maps further
clarifies the valley origin of the topology. In the $C=+1$ phase,
the smallest gap occurs near the $Y$ valley, indicating that the
dominant band inversion is rooted in the $Y$ sector. Conversely,
in the $C=-1$ phase, the inversion shifts to the $X$ valley. The
central rhombus–shaped region around $(M_{0},M_{1})\approx
(0,2t_{d})$ remains topologically trivial ($C=0$), even though the
spin–resolved valley gaps follow the shifted lines above. In this
regime, the two valleys contribute opposite Chern numbers in
opposite spin channels, yielding a vanishing total Chern number
but a finite composite spin Chern number $C_{s}=2$. We therefore
identify this region as a spin–valley Chern insulator.

\section{Conclusions}
We have shown that a $d$-wave pure altermagnet on a Lieb lattice provides a minimal platform for realizing magnetic-field-tunable quantum anomalous Hall phases with vanishing magnetization. The system hosts two inequivalent valleys at $X$ and $Y$, related by fourfold rotational symmetry. In the absence of a magnetic field, symmetry enforces opposite valley Chern numbers, leading to a phase diagram with a normal insulator, a spin Chern phase, and an intermediate topological semimetal characterized by a spin Chern number $C_s=2$. An external magnetic field breaks the rotational symmetry between valleys while preserving vanishing magnetization, allowing independent valley contributions. As a result, one valley can host $C=-1$ or $0$, and the other $C=0$ or $+1$, enabling globally nontrivial Chern phases. Berry-curvature analysis reveals robust gapped phases with $C=\pm1$ separated by valley-selective transitions. These results establish valleytronics as a route to topological phases in altermagnets without finite magnetization. 
Similar to the QAHE in ferromagnets that should be close to the QSH transition, when the altermagnet is close to the topological transition, on either the normal or spin Chern insulator side, the magnetic-field–induced transition to a Chern insulating phase occurs at relatively small fields. This enables rapid magnetic control of the quantum anomalous Hall effect and suggests applications in magnetization-free topological devices, while extensions to multilayer systems\cite{sun2025altermagnetizingfeseliketwodimensionalmaterials} could offer further control via thickness-dependent topology\cite{Safaei2015}.

\begin{acknowledgments}

The authors thank W. Brzezicki, S. Majewski and T. Dietl for useful discussions.
M.B.T. acknowledges the funding support by Narodowa Agencja Wymiany Akademickiej (NAWA) under the ULAM program with project number BPN/ULM/2025/1/00156/U/00001. M.B.T. acknowledges the funding support by Iran National Science Foundation (INSF) under project No.4043973.
This research was supported by the Foundation for Polish Science project “MagTop” no. FENG.02.01-IP.05-0028/23 co-financed by the European Union from the funds of Priority 2 of the European Funds for a Smart Economy Program 2021–2027 (FENG). We further acknowledge access to the computing facilities of the Interdisciplinary Center of Modeling at the University of Warsaw, Grant g91-1418, g91-1419, g96-1808, g96-1809 and g103-2540 for the availability of high-performance computing resources and support. We acknowledge the access to the computing facilities of the Poznan Supercomputing and Networking Center, Grants No. pl0267-01, pl0365-01 and pl0471-01.
\end{acknowledgments}

\medskip

\appendix

\bibliography{references}
\end{document}